\documentclass[fleqn,12pt]{article}
\usepackage{graphicx}
\usepackage{epsfig,color}

\definecolor{Blue}{named}{Blue}
\definecolor{Red}{named}{Red}
\definecolor{Green}{named}{ForestGreen}
\definecolor{Black}{named}{Black}
\definecolor{Olive}{named}{OliveGreen}
\definecolor{Lila}{named}{DarkOrchid}
\newcommand{\gsim}{\;\raisebox{-0.9ex}
           {$\textstyle\stackrel{\textstyle >}{\sim}$}\;}
\newcommand{\lsim}{\;\raisebox{-0.9ex}{$\textstyle\stackrel{\textstyle<}
           {\sim}$}\;}

\begin{document}
\begin{flushright}
  CERN-PH-TH/2006-128\\
  DCPT/06/96\\
  IPPP/06/48\\
  hep-ph/0607173
\end{flushright}

\begin{center}
{\Large\bf
Physics aspects of polarized $e^+$ at the
linear collider\footnote{Talk given at the Polarized Positrons Workshop `POSIPOL', CERN, 26--28 April 
2006}}\\[1em]
{\it Gudrid Moortgat-Pick} 
                                                                                
{\itshape TH Division, Physics Department, CERN, CH-1211 Geneva 23,
Switzerland}
\end{center}
                           
%

\begin{abstract}
Polarized $e^-$ and $e^+$ at the International Linear Collider play an important role
for discovering physics beyond the Standard Model and for precisely
unravelling the structure of the
underlying physics. The physics programme  
at the first energy stage at $\sqrt{s}=500$~GeV benefits strongly
from the polarization of both beams. 
But also at 1 TeV as well as
at a possible multi-TeV design of a linear collider, CLIC, the physics output is 
greatly enriched
by beam polarization.
An overview is given of the impact of providing polarized $e^+$ at the linear collider
in addition to polarized $e^-$ for physics 
studies in top, Higgs, supersymmetry and further models of physics 
beyond the Standard Model. 
\end{abstract}

\section{Overview}
\subsection{Physics programme at the ILC}
\begin{figure}
\begin{minipage}{10cm}
\caption{\label{fig_blue}
The study of Higgs particles will present a central part of the ILC physics programme.
Measurements of electroweak precision observables at LEP, SLD, CDF and D0 including the 
direct exclusion limit from LEP-2 predict that the mass of the Higgs in the SM 
$m_H \le 207$ {\rm GeV}\quad (95\%\hspace{.2cm}{\rm CL})~\cite{lepewwg}.
In supersymmetric theories the limit is at about $m_h\sim 135$~GeV~\cite{georg-higgs}.}
\end{minipage}
\begin{minipage}{5cm}
\setlength{\unitlength}{1cm}
\begin{picture}(5,5)
\setlength{\unitlength}{1cm}
\put(0,0){\mbox{\includegraphics[height=.22\textheight,width=0.33\textheight]{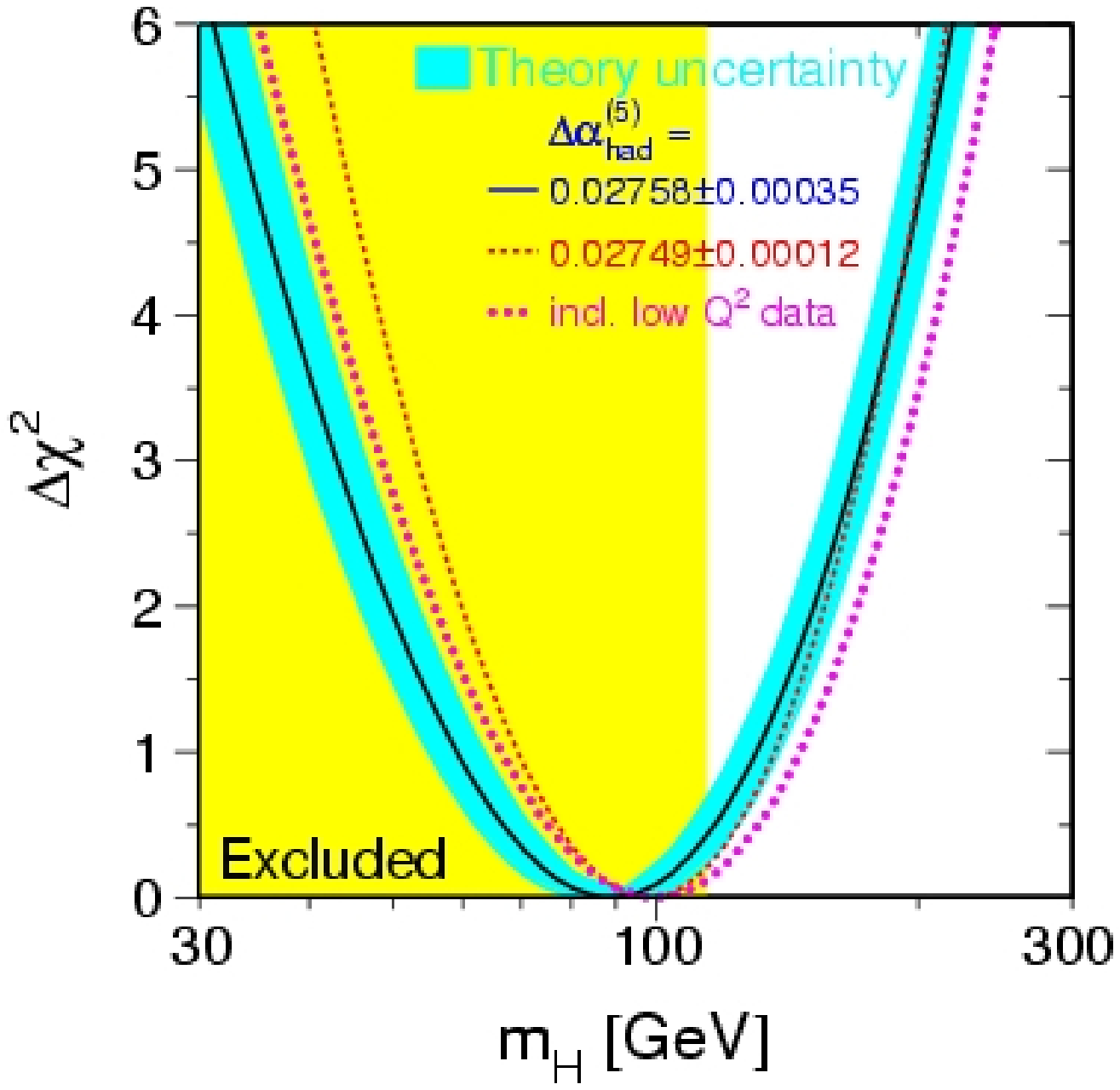}}}
\end{picture}
\end{minipage}
\end{figure}

The International Linear Collider (ILC) will start with a first energy
phase of $\sqrt{s}\le 500$~GeV, which is perfectly suited to precision
top and Higgs studies.
Precise measurements of the properties of the top
quark, which is by far the heaviest known elementary particle, will
greatly advance our understanding of the underlying physics at the
quantum level~\cite{georg-nature}.  
Electroweak precision data indicate in the Standard
Model (SM) a light Higgs with a mass below about 207~GeV 
(see Fig.~\ref{fig_blue}). It will be
crucial to precisely determine the mass, couplings, spin and
CP properties of the new particle in order to experimentally establish
the mechanism of electroweak symmetry breaking.

A further highlight of expected physics at the ILC will be the discovery and determination of 
new physics beyond the SM. The most prominent candidate for new physics 
is supersymmetry (SUSY). Fits with electroweak precision data and experimental bounds from
collider and cosmological experiments are consistent with light SUSY, indicating that
the energy range of $\sqrt{s}=500$--$1000$~GeV will be perfectly suited to 
the discovery and the precise measurements of the properties of SUSY particles, at 
least their light spectrum. These results, 
together with results from the Large Hadron Collider (LHC),  will allow us to unravel the 
underlying structure of the theory and to predict the properties of expected 
heavy SUSY particles. 

Some new physics scales
could be too large to be directly accessible at the LHC or at the ILC.
The ILC also has a large discovery potential,
complementary to that of the LHC, for indirect searches of physics
beyond the kinematic limit. Manifestations of such new interactions
can be probed through deviations of cross sections from the SM
predictions, and indirect bounds on the new energy scales and coupling
constants can thereby be derived.

The results from the ILC, together with those of the LHC,
will outline the needed requirements and energy scale 
for a multi-TeV linear collider, CLIC, whose feasibility studies are currently 
being designed.

\subsection{Polarized beams at linear colliders}
{\bf a) History and future}\\
The use of polarized beams plays an important role in the whole physics programme of a
linear collider. A prominent example from history to demonstrate the importance of 
beam polarization is the measurement of the electroweak 
mixing angle at the SLC (the SLAC Linear Collider) with a precision of
$\Delta \sin\theta_{\rm eff}=0.00026$~\cite{lepewwg}. 
At LEP, in spite of the very high luminosity but
without polarization of the beams for the physics runs, a precision of  
$\Delta \sin\theta_{\rm eff}=0.00029$~\cite{lepewwg} was reached. 

The polarization of the electron beam is foreseen for the baseline
design of the ILC~\cite{BCD}. A high degree of at least 80\% longitudinal polarization is
envisaged, but new results indicate that even 90\% should be
achievable. Two different sources for the production of polarized $e^+$ 
are currently under discussion, either via undulator radiation 
(preferred solution for the ILC facility) 
or via laser-backscattering  processes (preferred solution for the CLIC design), 
both leading to a longitudinal polarization degree 
of about 60\% for the $e^+$ beam without any loss in luminosity. Higher $e^+$ polarization
of about 75\%  
is possible at a cost in luminosity. With spin rotators
the longitudinal polarization can be rotated to provide also transversely polarized 
beams for physics studies.\\

\noindent {\bf b) Introductory remarks and definitions}\\
In~\cite{Moortgat-Pick:2005cw}, it is shown that
the full potential of the ILC could be realized only with
the polarization of both the $e^-$ and $e^+$ beams. 
Polarized $e^+$ serve either as i) a substantial factor for the physics results 
or/and also as ii) a rather easily obtainable lucrative factor. Both i) and ii) are important 
to optimize the physics outcome.

\begin{figure}[htb]
\begin{minipage}{14cm}
\setlength{\unitlength}{1cm}
\begin{picture}(6,5.5)
\setlength{\unitlength}{1cm}
\put(0.3,5.2){\small $P_{\rm eff} [\%]$}
\put(7,.2){\small $P_{e^+} [\%]$}
\put(2.2,4.8){\small\color{Blue} $P_{e^-}=-90\%$}
\put(2.8,3.7){\small\color{Olive} $P_{e^-}=-80\%$}
\put(4.5,3){\small\color{Red} $P_{e^-}=-70\%$}
\put(1,0.5){\epsfig{file=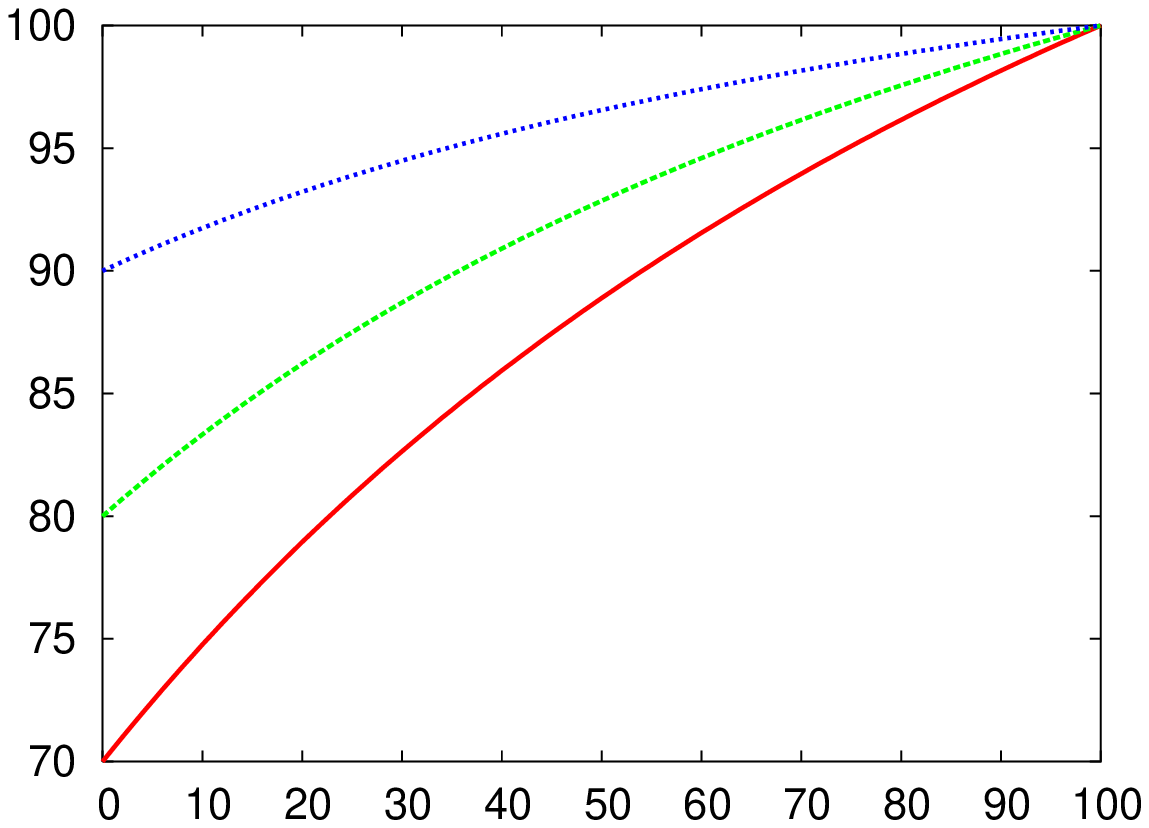,scale=.58}}
\put(9.1,5.2){\small $\frac{1}{x}\frac{\Delta P_{\rm eff}}{|P_{\rm eff}|}$}
\put(9.1,4.5){\small $\sim \frac{\Delta A_{\rm LR}}{A_{\rm LR}}$}
\put(16,0.2){\small $P_{e^+} [\%]$}
\put(13.,1.4){\small\color{Blue} $P_{e^-}=-90\%$}
\put(14.7,2.3){\small\color{Olive} $P_{e^-}=-80\%$}
\put(13.7,2.8){\small\color{Red} $P_{e^-}=-70\%$}
\put(10.,0.5){\epsfig{file=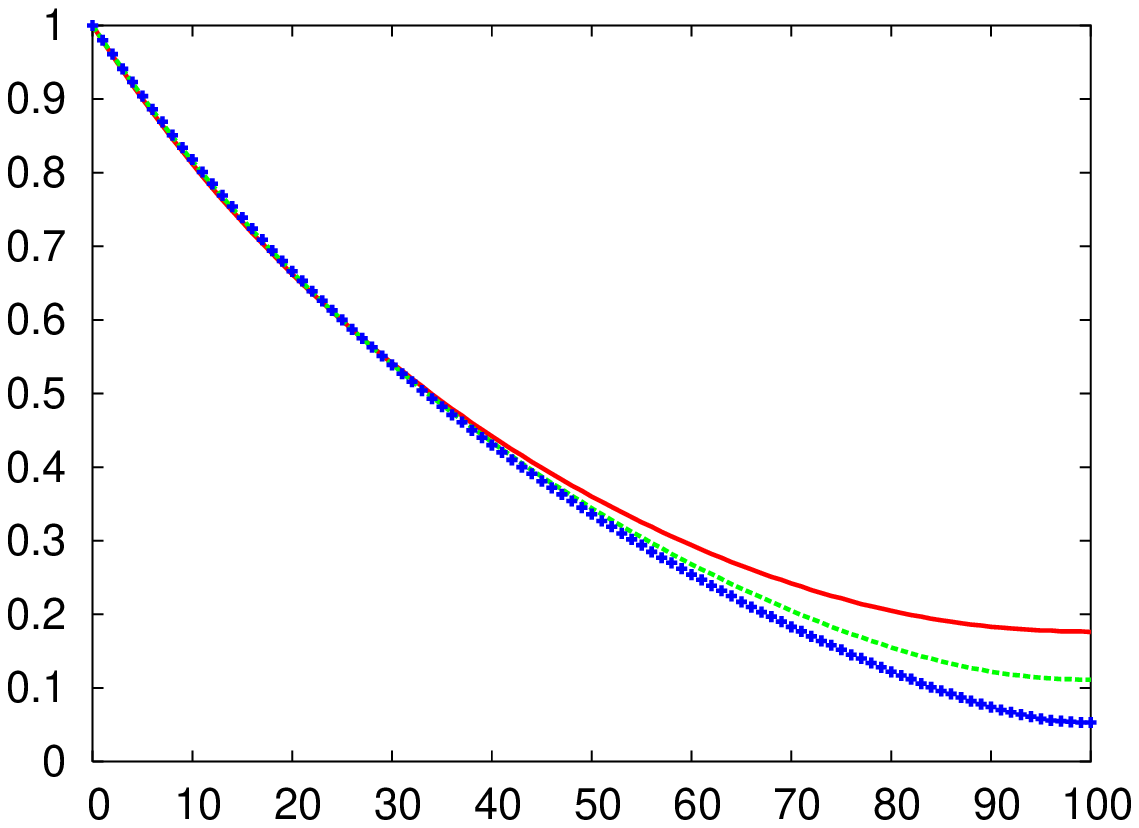,scale=.58}}
\end{picture}
\end{minipage}
\begin{minipage}{17.5cm}
\caption{ Left: effective polarization as a function of positron polarization. 
Right: relative uncertainty on the effective polarization 
$\Delta P_{\rm eff}/|P_{\rm eff}|\sim \Delta A_{\rm LR}/A_{\rm LR}$, normalized to the 
relative polarimeter precision $x=\Delta P_{e^-}/P_{e^-}=\Delta P_{e^+}/P_{e^+}$.
\label{fig_peff}}
\end{minipage}
\end{figure}

Physics processes occur through $e^-e^+$ annihilation (`$s$-channel diagrams') and 
scattering (`$t$, $u$-channel diagrams'). In annihilation diagrams the helicities of the incoming
beams are coupled to each other, whereas in scattering processes, they
are coupled to those of the final particles and therefore are directly sensitive to their 
chiral properties. In processes where only (axial-) vector interactions are contributing,
the cross section with polarized beams is given by:
\begin{equation}
 \sigma(P_{e^-} P_{e^+})= (1-P_{e^-} P_{e^+}) \sigma_{\rm unpol} [1-P_{\rm eff} A_{\rm LR}],
\label{eq_peff}
\end{equation}
where $A_{\rm LR}$ denotes the left--right asymmetry and $P_{\rm eff}$ 
the effective polarization, given by $P_{\rm eff}=[P_{e^-}-P_{e^+}]/[1-P_{e^-}P_{e^+}]$.
Polarized $e^+$ lead to the improvement of
the effective polarization and enhance the precision in measurements of the
left--right asymmetry (see Fig.~\ref{fig_peff}), which are, for instance, often exploited in the
high-precision studies of the SM. 

In addition to the enhancement in the effective polarization, which is important for 
many precision studies in the SM,
the polarization of both beams
enables detailed analyses of the properties of new particles and of
new kinds of interactions, as well as of indirect searches with high
sensitivity for new physics in a widely model-independent approach. In the following
an overview is given about some physics examples; more details and more examples
can be found in \cite{Moortgat-Pick:2005cw}.

\section{Polarized beams in top searches}
{\bf a) Determination of electroweak properties}\\
A linear collider provides an ideal tool to probe the couplings of the
top quark to the electroweak gauge bosons. The neutral electroweak
couplings are accessible only at lepton colliders, because top quarks
at hadron colliders are pair-produced by gluon exchange.
 
The most general $(\gamma,Z)t \bar{t}$ couplings can be written as
\begin{equation}
\Gamma^{\mu}_{t\bar{t}\gamma,Z}=i e
\left\{\gamma^{\mu}[F^{\gamma,Z}_{1V}
+F^{\gamma,Z}_{1A}\gamma^5]+\frac{(p_t-p_{\bar{t}})^{\mu}}{2
m_t}[F^{\gamma,Z}_{2V} +F^{\gamma,Z}_{2 A}\gamma^5]\right\},
\label{eq_topform}
\end{equation}
where $F^{\gamma}_{1 V}$, $F^{Z}_{1V}$,
$F^Z_{1A}$ denote the only form factors that are different from zero in the SM. 

Polarization effects have been studied at the top threshold \cite{Kuehn}.
In the SM the main production process occurs via $\gamma$, $Z$ exchange.
To determine the SM top vector coupling $v_t$, one has to measure the left--right
asymmetry $A_{\rm LR}$ with high accuracy. With an integrated luminosity of
${\cal L}_{\rm int}=300$~fb$^{-1}$, precisions in $A_{\rm LR}$ and $v_t$ of about $0.4\%$
and $1\%$, respectively, can be achieved at the ILC. The gain in using
simultaneously polarized $e^-$ and $e^+$ beams with
$(P_{e^-},P_{e^+})=(\mp80\%,\pm60\%)$
is given by the higher effective polarization of $P_{\rm eff}=95\%$ compared to the
case of only polarized electrons with $|P_{e^-}|=80\%$. This leads,
according to Fig.~\ref{fig_peff}, to a reduction of the relative uncertainty 
$\Delta A_{\rm LR}/A_{\rm LR}\simeq \Delta P_{\rm eff}/ P_{\rm eff}$ by a factor of about 3.

Limits to all the above mentioned form factors have also been derived
in the continuum at $\sqrt{s}=500$~GeV for unpolarized beams and
$(|P_{e^-}|,|P_{e^+}|)=(80\%,0)$.
It has been estimated that the polarization of both beams with
$(|P_{e^-}|,|P_{e^+}|)=(80\%,60\%)$ leads to an increase of the $t\bar{t}$ cross section
by a factor of about $\sim 1.5$ and improves again the bounds by about a
factor 3~\cite{Moortgat-Pick:2005cw}. Complete simulations are still missing.\\

\noindent {\bf b) Limits for CP- and FCN-violating couplings}\\
Searches for
anomalous $t{\bar t}\gamma$ and $t{\bar t} Z$ couplings can be made by
studying the decay energy and angular distributions of $l^+$ ($l^-$) or
$b$ ($\bar b$) in $e^+e^-\to t{\bar t}$ followed by the subsequent decays
$t\to l^+\nu_l b$
(${\bar t}\to l^- {\bar \nu}_l {\bar b}$).
Focusing on CP violation, a suitable observable is represented by the
forward--backward charge asymmetry \cite{Grzadkowski:2000nx,Rindani:2003av}.
With $\sqrt s=500\hskip 3pt {\rm GeV}$,
${\cal L}_{\rm int}=500\hskip 3pt{\rm fb}^{-1}$, and 60\% reconstruction
efficiency for either lepton or $b$, the forward--backward charge asymmetry
could be measured at the 5.1$\sigma$ (2.4$\sigma$) level for $b$-quarks (leptons)
assuming CP-violating couplings of the order of $5\times 10^{-2}$ and
unpolarized beams. Having both beams 80\% polarized, the reach on
${\cal A}_{\rm CP}^f$ would even increase up to 16$\sigma$ (3.5$\sigma$).

Flavour-changing neutral (FCN) couplings of the top quark are relevant to
numerous extensions of the SM, and can represent an interesting field for
new-physics searches. Limits on top FCN decay branching ratios can be obtained
from top-pair production, with subsequent $\bar t$ decay into $\gamma, Z$ plus
light quark governed by the FCN anomalous $tVq$ couplings
($V=\gamma, Z$ and $q=u,c$), $e^+e^-\to t{\bar t}\to W^+ b V {\bar q}$,
or from single top production
$e^+e^-\to t{\bar q}\to W^+ b {\bar q}$ mediated by the anomalous
couplings at the production vertex.
Single-top production is more sensitive to top anomalous couplings, but
top decays help to disentangle the type of anomalous coupling
involved.
Beam polarization is
very efficient in significantly reducing the background and is therefore particularly important
in limits obtained from single-top production.
 The background is essentially
dominated by the $W^+ +{\rm 2 jets}$ final state, with $W^+$ decaying into
$l \nu$ and one jet misidentified as a $b$-jet.

With polarization $(80\%,0)$, the background decreases by a factor of
$1/(1-P_{e^-})\approx5$ while keeping 90\% of the signal.  With $(80\%,-45\%)$
the background is reduced by a factor of
$1/(1-P_{e^-})(1+P_{e^+})\approx9$ and the signal is increased by 20\%
with respect to the case of no polarization \cite{Aguilar-Saavedra:2001ab}.
In conclusion, $S/B$ and $S/\sqrt{B}$ are improved by factors of 2.1 and 1.7,
respectively.
Already with $e^-$ and $e^+$ polarization (80\%, 45\%), as an example, the 3$\sigma$ 
discovery limits on the
vector ($\gamma^{\mu}$) coupling at
$\sqrt{s}=500$~GeV is improved by a factor of 3 (a factor of 1.7 with respect to only electron
polarization) and the limits on the tensor ($\sigma^{\mu\nu}$) coupling at
$\sqrt{s}=800$~GeV by a factor of about 2.6 (a factor 1.8 with respect to electron
polarization only).\\

\noindent {\bf c) Transversely-polarized beams in top studies}\\
The observation of
CP violation in $e^+e^-$ collisions requires either the measurement of the polarization
of the final-state particles or the availability of polarized
beams. Transverse polarization of initial beams defines one more direction,
and can provide CP-odd asymmetries without having to measure
final-state polarizations directly. This may represent an advantage, e.g.\ as regards
the statistical significance of the signal. Both beams, however, have to be polarized, 
otherwise all effects at the leading order from transverse polarization vanish 
for $m_e \to 0$ (suppression by $m_e/\sqrt{s}$).

In e.g.\ $e^+e^- \to t \bar{t}$ production, only (pseudo-)
scalar or tensor currents associated with a new-physics scale
can lead to CP-odd observables at the leading order
in the new interaction, if transversely-polarized beams are used.
They are due to the interference
 between these new currents and the $\gamma$ and $Z$ exchanges in
the $s$-channel. These interference terms
cannot be seen with longitudinally-polarized or unpolarized beams: both beams, $e^-$ and $e^+$,
have to be transversely polarized.
The corresponding new-physics
scale $\Lambda$ can be bounded at the 90\% confidence level, at about $7$~TeV,
with $\sqrt{s}=500$~GeV and $(P_{e^-},P_{e^+})=(80\%,60\%)$; 
see~\cite{Ananthanarayan:2003wi,Moortgat-Pick:2005cw} for details.

\section{Polarized beams in Higgs analyses}
A striking and unique goal of physics at a linear collider will be the clear establishment 
of the mechanism of the electroweak symmetry breaking, which requires the precise 
measurement of all Higgs couplings.
Beam polarization does play an important lucrative factor
in determining the Higgs properties at energies $\sqrt{s}\le 500$~GeV.\\[-.7em]

\noindent {\bf a) Separation of the production processes}\\
Assuming a light SM Higgs with $m_H\le 130$~GeV, which is the range
preferred by both fits of precision observables in the SM
\cite{Delphi} and predictions of SUSY theories (see
e.g. \cite{georg-higgs}), Higgs-strahlung dominates for $\sqrt{s}\lsim
500$~GeV and $WW$ fusion for $\sqrt{s}\gsim 500$~GeV.  At a LC with
$\sqrt{s}=500$~GeV and unpolarized beams, the two processes have
comparable cross sections.
Beam polarization can be used to enhance the $HZ$ contribution with respect to the $WW$
fusion signal, and vice versa, and to suppress the dominant SM background of 
$WW$ production significantly.
Table~\ref{back_WW} shows that there is a gain of a factor
$(1.26/0.08)/(0.87/0.20)\sim 4$ in the ratio
$\sigma(HZ)/\sigma(H\nu\bar{\nu})$ when left-handed polarized positrons are
used in addition to right-handed polarized electrons.\\[-.7em]

\noindent {\bf b) Determination of general Higgs couplings}\\
Using an optimal-observable method, which allows
the minimization of statistical uncertainties on the
couplings, we can reach a high accuracy in the determination of the general $ZZH$
and $Z\gamma H$ couplings. In~\cite{Hagiwara:2000tk} it was shown that
beam polarization is essential for determining the sensitivity to the
seven general couplings. Simultaneous
polarization of the $e^+$ and $e^-$ beams results in an increase in
the sensitivity, so that for $\sqrt{s}=500$~GeV, ${\cal L}_{\rm
int}=300$~fb$^{-1}$ and $(P_{e^-},P_{e^+})=(\pm 80\%,60\%)$, the sensitivity is
improved by about 30\% with respect to the case of $(\pm 80\%,0)$.\\[-.7em]

\noindent{\bf c) Measurement of the top Yukawa couplings}\\
By virtue of its large mass, the top quark has the largest Yukawa coupling
to the Higgs boson of all fermions: $g_{ttH}\simeq 0.7$, to be compared for
instance with $g_{bbH}\simeq 0.02$. It plays a key role in the
mechanism of electroweak symmetry breaking and mass generation.
Therefore, an accurate measurement of the top--Higgs Yukawa coupling is
particularly important. At the LHC a determination of the Yukawa coupling with a precision
of about 20\% is expected; however, some model assumptions have to be made.

At the linear collider the process $e^+e^- \to t\bar{t}H$
provides the best opportunity for a direct and precise determination
of the top--Higgs Yukawa coupling through the cross-section measurement. This measurement is
particularly challenging because of the smallness of the $t\bar{t}H$ cross section,
e.g. $\sigma_{ttH}\simeq 2.5$ fb for $m_H=120$ GeV at $\sqrt{s}=800$ GeV,
and of the very large background, dominated by $t\bar{t}$+jets. 
Since, in the Baseline Configuration Document (BCD) for the ILC~\cite{BCD}, 
$\sqrt{s}=500$~GeV is chosen for the first stage of the ILC, it is
important to assess the feasibility of measuring the top Yukawa coupling at this energy.
At $\sqrt{s}=500$ GeV, the $t\bar{t}H$ cross section is reduced by
a factor of about 10 with respect to $\sqrt{s}=800$ GeV, e.g.
$\sigma_{ttH}\simeq 0.2$ fb for $m_H=120$ GeV.

 A preliminary estimate~\cite{juste2} yields $\Delta
g_{ttH}/g_{ttH}\simeq 24\%$ for $m_H=120$ GeV, assuming ${\cal
L}_{\rm int}=1000$ fb$^{-1}$ but unpolarized beams.  
The usage of beam polarization 
plays an important role by increasing the cross section. 
As shown in eq.~(\ref{eq_peff}), there are two
enhancement factors that can be exploited: $(1-P_{e^+}P_{e^-})$ and
$[1-P_{\rm eff}A_{\rm LR}]$ (in the case of SM $t\bar{t}H$ production,
$A_{\rm LR}\simeq +0.44$).  A recent study~\cite{juste3} at
$\sqrt{s}=500$ GeV has shown that for 
$(P_{e^-},P_{e^+})=(-0.8,+0.6)$ the $t\bar{t}H$ cross section can be
increased by a factor of about $\sim 2.1$, resulting in an improvement
in the precision of $g_{ttH}$ of $45\%$. In the case where no positron
polarization is available, $(P_{e^-},P_{e^+})=(-0.8,0)$, the
improvement would only be $19\%$.  The precision in measuring the Yukawa
coupling $g_{ttH}$ can be improved by a factor of about 2.5 with respect to
the case of having only polarized electrons. Polarized $e^+$ can
therefore play a substantial role in the determination of the top
Yukawa couplings at the first stage of the ILC, also improving the 
predicted precision at the LHC significantly.\\ 

\noindent{\bf d) Measurement of the triple Higgs couplings}\\
The determination of the Higgs potential is crucial to establish the Higgs mechanism.
Therefore also the triple Higgs couplings have to be precisely measured, either in 
the $HHZ$ or in the $HH\nu\bar{\nu}$ final state. At $\sqrt{s}=500$~GeV the triple 
Higgs couplings can only be determined in the $HHZ$ channel; former studies~\cite{triplehiggs,tdr} 
with 
unpolarized beams predict a precision of 22\%. 

To determine the triple Higgs couplings in the $HH\nu\bar{\nu}$ final state, higher energies 
are definitely needed. At 3 TeV at CLIC the $WW$ fusion process is kinematically enhanced and 
a precision of up to about 13\%~\cite{CLIC} for the determination of the triple 
Higgs coupling is predicted.

However, both studies have been done only for unpolarized beams. Complete simulations 
with polarized beams and realistic detector simulations and background suppression 
are still missing. However, it is estimated that at least
a further increase of about 50\% is expected. Regarding the moderate precision at 
$\sqrt{s}=500$~GeV,  using both beams polarized can be a rather 
substantial improvement factor 
for the obtainable precision in the triple Higgs coupling.

\begin{table}[htb]
\begin{center}
{\small
\begin{tabular}{|c||c|c||c|c|}
\hline Beam polarization & $e^+e^- \to H\nu \bar{\nu}$ & $e^+e^- \to HZ$ & 
$e^+e^-\to W^+W^-$ & $e^+e^- \to ZZ$ \\\hline 
$(+80\%, 0)$ & 0.20 & 0.87 & 0.20 & 0.76\\ \hline 
$(-80\%, 0)$ & 1.80 & 1.13 & 1.80 & 1.25\\ \hline\hline 
$(+80\%, -60\%)$ & 0.08 & 1.26 & 0.10 & 1.05\\\hline 
$(-80\%,+60\%)$ & 2.88 & 1.70 & 2.85 & 1.91 \\ \hline
\end{tabular}
\caption{ Scaling
factors of Higgs production, in Higgs-strahlung and $WW$ fusion, and of the dominant SM 
background processes $WW$ and $ZZ$ production, at 
$\sqrt{s}=500$~GeV, for several polarization configurations
compared with the unpolarized
case~\cite{Desch_obernai,Moortgat-Pick:2001kg}.
\label{back_WW}
}}
\end{center}
\end{table}

\vspace{-1.2cm}
\section{Polarized beams in searches for supersymmetry}
One of the most promising candidates for physics beyond the SM is supersymmetry.
SUSY can solve open problems of the SM as, for instance, the hierarchy problem; it 
enables gauge unification and provides candidates for cold dark matter. 
Furthermore SUSY models have high predictive power, and precise calculations for future
experiments can be made.
This new symmetry predicts that every SM particle has a SUSY partner that has the same
quantum numbers as their SM partner, with the exception of the spin. To really establish
supersymmetry experimentally, all model assumptions and implications 
have to be verified. Furthermore the fundamental underlying parameters have to determined precisely.
Since the number of new parameters is large, even in the Minimal 
Supersymmetric Standard Model (MSSM), there are 105, this task 
may be very challenging.

\begin{figure}[htb]
\begin{minipage}{6cm}
\setlength{\unitlength}{1cm}
\begin{picture}(5,5)
\setlength{\unitlength}{1cm}
\put(0,0){\mbox{\includegraphics[height=.22\textheight,width=0.33\textheight]{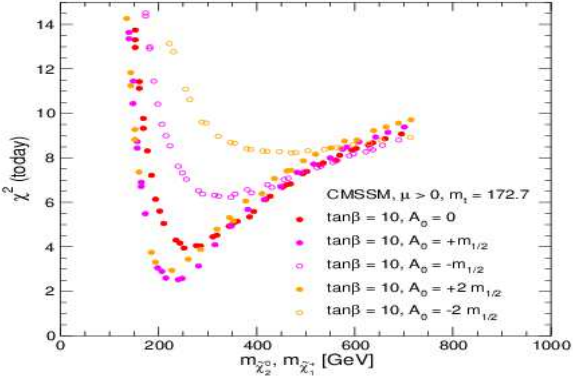}}}
\end{picture}
\end{minipage}\hspace{2.5cm}
\begin{minipage}{9cm}
\caption{The $\chi^2$ function for the electroweak observables
$M_W$, $\sin^2\theta_{\rm eff}$, $(g-2)_{\mu}$, BR($b\to s\gamma$) and
$M_h$ evaluated in the CMSSM for $\tan\beta=10$ and $m_t=172.7\pm
2.9$~GeV, matching the central value of the relic neutralino density
indicated by WMAP~\cite{weiglein}. \label{fig_susyfit}}
\end{minipage}
\end{figure}

One crucial question is how large the scale of this new symmetry is predicted to be.
 In order to be consistent with electroweak precision measurements and cosmological bounds,
at least some of the electroweak interacting SUSY particles are predicted to be rather light
and should be accessible at the ILC with $\sqrt{s}=500$~GeV. In a recent 
study~\cite{weiglein}
a parameter fit was applied within 
the Constrained MSSM (CMSSM), leading
to the prediction of, for instance, light neutralinos
$\tilde{\chi}^0_{1,2}$ and a light chargino $\tilde{\chi}^{\pm}_1$, 
so that the processes $\tilde{\chi}^0_1\tilde{\chi}^0_2$ and 
$\tilde{\chi}^+_1\tilde{\chi}^-_1$ should be accessible at the ILC (Fig.~\ref{fig_susyfit}). \\

\noindent {\bf a) Parameter determination}\\
In \cite{ckmz} it has been
demonstrated that already these light pairs are sufficient to determine the fundamental 
SUSY parameters and to predict the heavier SUSY particles. Polarized $e^+$ lead 
to an essential lucrative factor and can become rather substantial to suppress 
background processes. Followed by combined analyses at the LHC
and the ILC~\cite{lhclc}, a promising framework to unravel precisely the structure of the 
SUSY model is provided, even if 
the complete SUSY particle spectrum is not accessible.
 Polarized $e^-$ and $e^+$ beams are crucial in that context,
not only to enhance the cross section and to suppress background processes, but also 
to provide more observables, which are essential for determining the parameters with as few  
model assumptions as possible. This fact is  
even more important in cases where only a part of the particle spectrum is accessible,
for instance at the first energy stage of the ILC with $\sqrt{s}=500$~GeV.
\\

\begin{figure}[htb]
\begin{minipage}{10cm}
\hspace{1cm}
\setlength{\unitlength}{1cm}
\begin{picture}(6,6.5)
\setlength{\unitlength}{1cm}
\put(-1.1,3){\mbox{\includegraphics[height=.14\textheight]{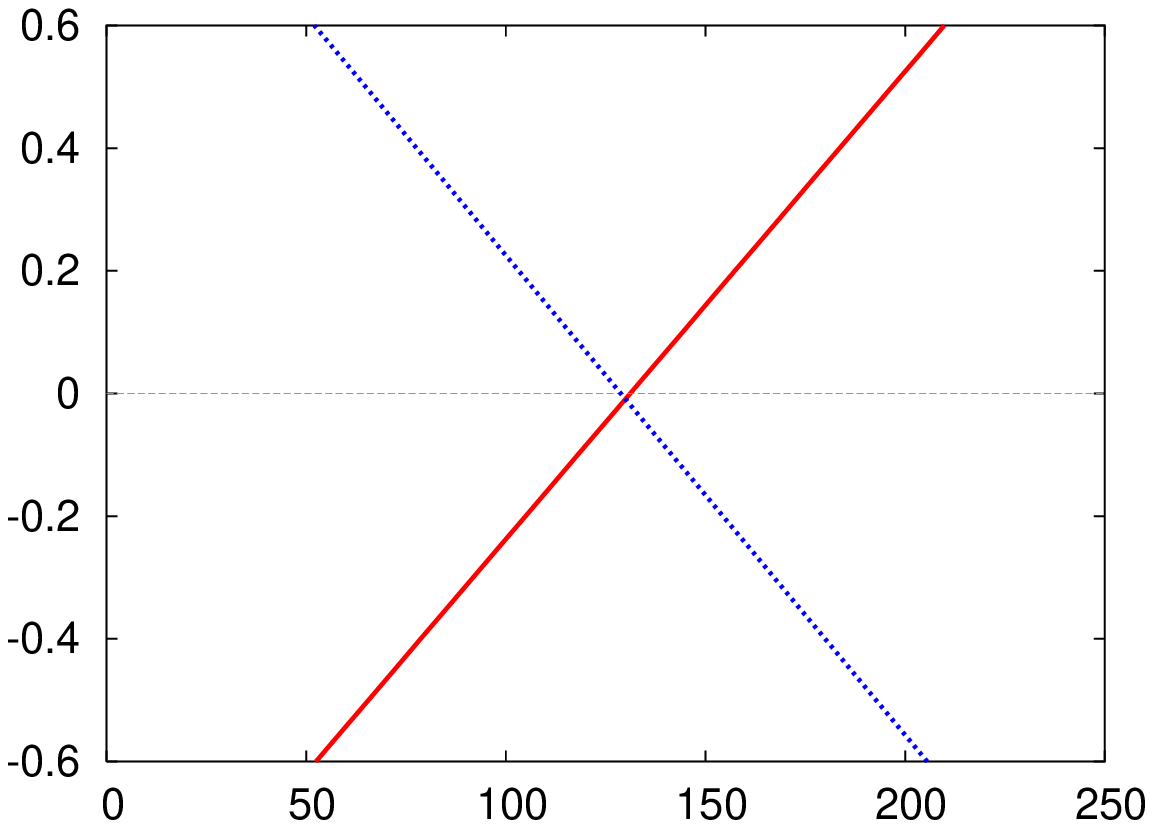}}}
\put(-.6,3.){\mbox{\includegraphics[height=.015\textheight,width=.22\textheight]{0028-box}}}
\put(-1.1,0.1){\mbox{\includegraphics[height=.14\textheight]{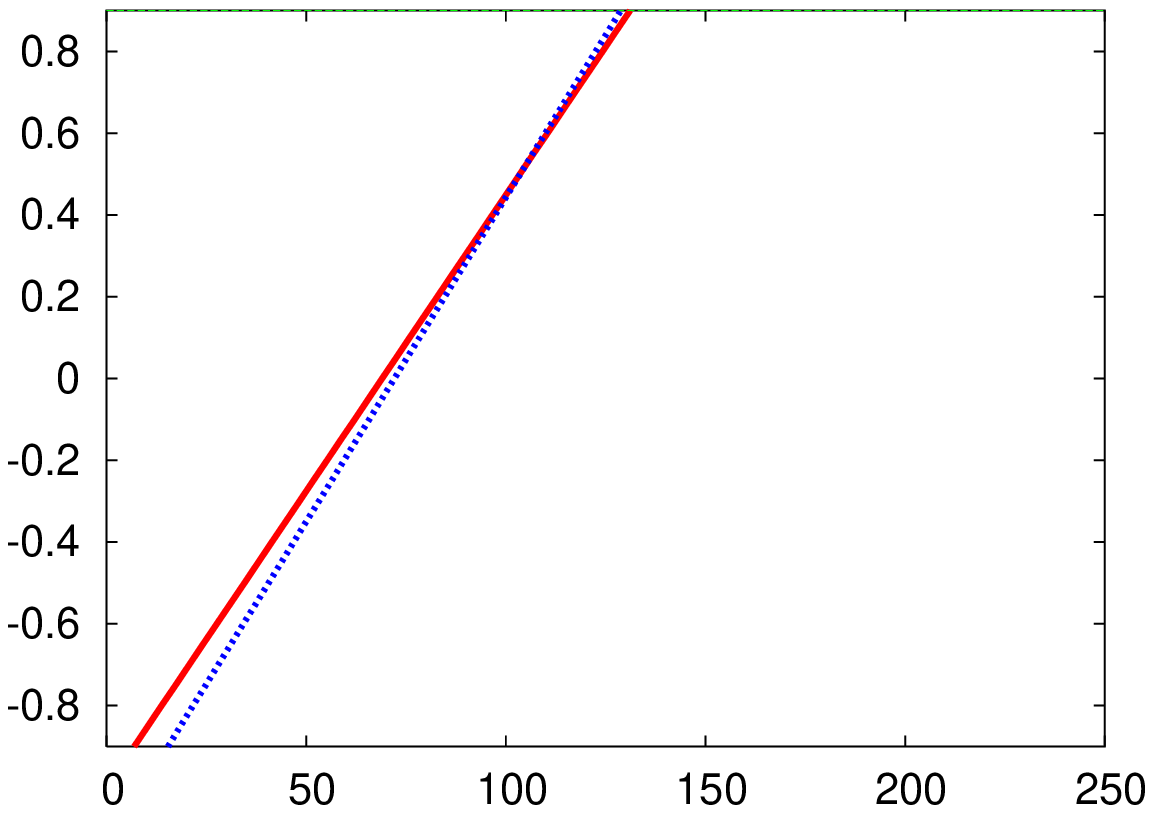}}}
\put(-1.25,4.7){\footnotesize $P_{e^+}$}
\put(-1.25,1.7){\footnotesize $P_{e^-}$}
\put(-1.4,3.1){\tiny $0.9\to$}
\put(.8,5.8){\scriptsize $P_{e^-}=+0.9$}
\put(2.4,5.4){\scriptsize\color{Red} $\tilde{e}^+_L\tilde{e}^-_R$}
\put(.0,5.4){\scriptsize\color{Blue} $\tilde{e}^+_R\tilde{e}^-_R$}
\put(1.7,2.5){\scriptsize\color{Red}$\tilde{e}^+_L\tilde{e}^-_R$}
\put(.1,2.5){\scriptsize\color{Blue} $\tilde{e}^+_R\tilde{e}^-_R$}
\put(1.43,3.7){\Large$\uparrow$}
\put(1.4,.5){\scriptsize $\sqrt{s}=500$~GeV}
\put(-.4,6.4){\scriptsize
$e^+e^-\to\tilde{e}^+_{L,R}\tilde{e}^-_{R}\to e^+e^- \tilde{\chi}^0_1\tilde{\chi}^0_1$}
\put(3.5,3.1){\includegraphics[width=.2\textheight,height=0.13\textheight]{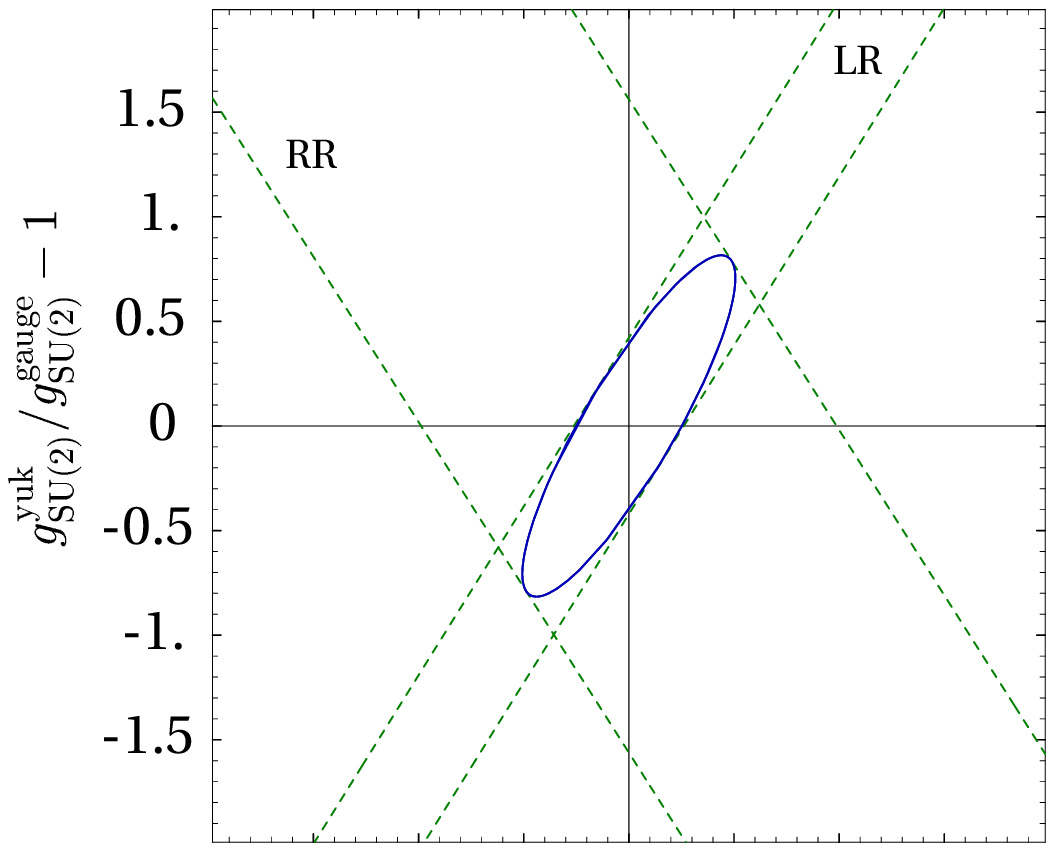}}
\put(3.5,0.15){\includegraphics[width=.2\textheight,height=0.13\textheight]{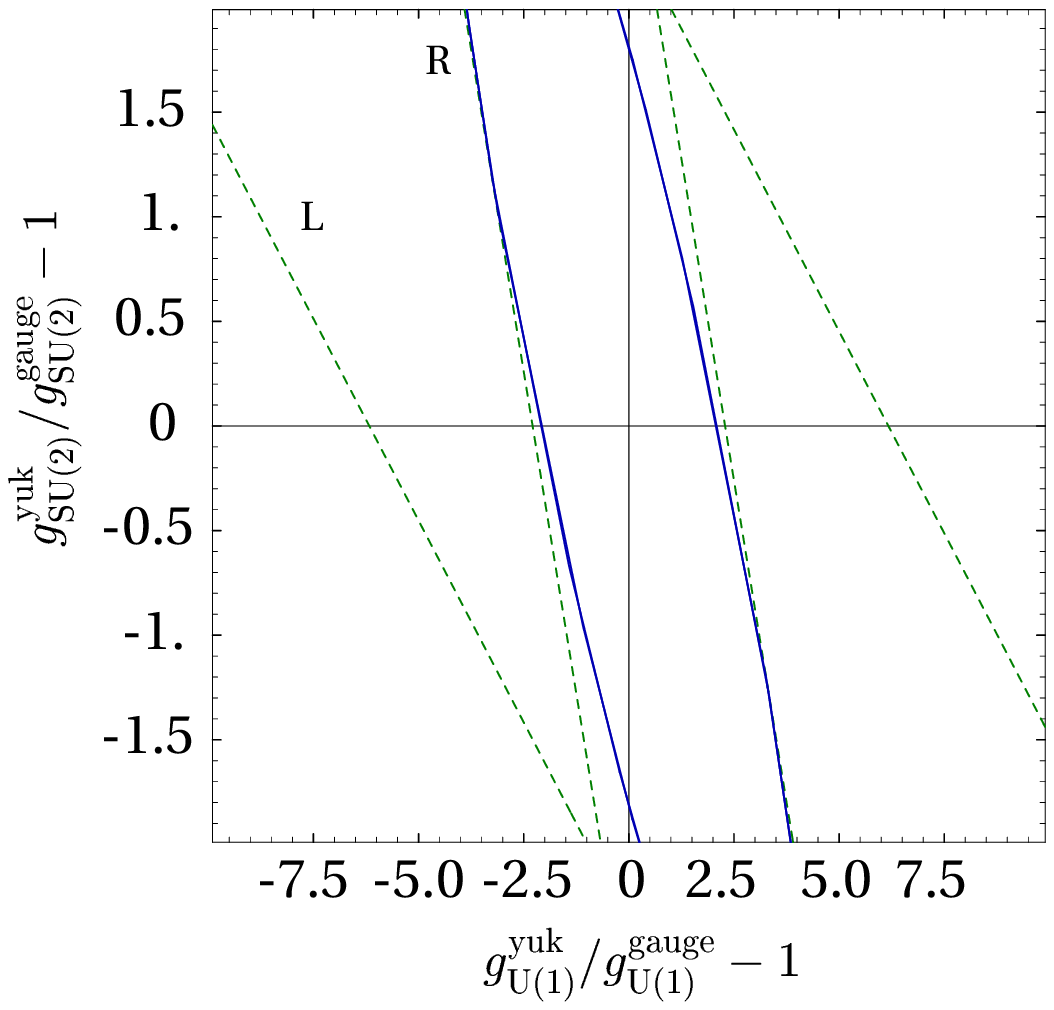}}
\put(4.2,6.4){\scriptsize
$e^+e^-\to\tilde{e}^+_{L,R}\tilde{e}^-_{R}\to e^+e^- \tilde{\chi}^0_1\tilde{\chi}^0_1$}
\put(.5,-.2){\scriptsize cross section [fb]}
\end{picture}
\end{minipage}
\begin{minipage}{7.5cm}
\vspace{-.5cm}
\caption{\label{fig_susyqn} Left: Test of chiral quantum numbers -- 
separation of the selectron pair
$\tilde{e}_{\rm L}^+\tilde{e}^-_{\rm R}$ is not possible
with $e^-$ polarization alone, $P_{e^-}=+90\%$
 (lower plot). If both beams are polarized,
the RR configuration separates the pairs $\tilde{e}_{\rm L}^+\tilde{e}^-_{\rm R}$,
 $\tilde{e}_{\rm L}^+\tilde{e}^-_{\rm L}$
(see arrow, upper plot)~\cite{Moortgat-Pick:2005cw}.
 Right: Test of Yukawa couplings -- 
 1$\sigma$ bounds on the determination of the U(1) and 
SU(2) Yukawa couplings between $e^+$,
$\tilde{e}_{\rm R,L}^+$, $\tilde{\chi}^0_i$; R (L) means $P_{e^-} = +90\%$ ($-90\%$) (lower
plot) and RR, LR means  $(P_{e^+},P_{e^-}) = (+60\%,+90\%)$, $(-60\%,+90\%)$ (upper plot).
Both studies are done at $\sqrt{s}=500$~GeV~\cite{Moortgat-Pick:2005cw}.}
\end{minipage}
\end{figure}

\noindent {\bf b) Tests of the quantum numbers in the scalar particle sector}\\
Prominent examples of the scalar SUSY sector are the
selectrons and spositrons $\tilde{e}^{\pm}_{\rm L,R}$,
which have to be associated to their chiral SM partners, the left- and
right-chiral electrons and positrons.
This association can be directly tested in the production
of the pairs {\color{Red}$\tilde{e}^+_{\rm L}\tilde{e}^-_{\rm R}$}
produced only in the $t$-channel process. The process
must be experimentally separated from the pair
{\color{Blue}$\tilde{e}^+_{\rm R}\tilde{e}^-_{\rm R}$} produced also in
the $s$-channel. It has been shown that even a
highly polarized electron beam will not be sufficient to separate the pairs,
since both can be produced with almost identical cross sections and have the same decay; 
see Fig.~\ref{fig_susyqn} (left lower plot).
Applying simultaneously polarized positrons, the pairs get
different cross sections,
can be isolated, and the $\tilde{e}_{\rm L}^+$ and $\tilde{e}^-_{\rm R}$ can be
identified by charge separation;
see Fig.~\ref{fig_susyqn} (left upper plot)~\cite{Blochinger:2002zw,Moortgat-Pick:2005cw}.

As another consequence of SUSY, the SU(2) and U(1) SUSY Yukawa
couplings have to be identical to the corresponding SM gauge
couplings.  Assuming that the masses and mixing parameters of the
neutralinos have been predetermined in the gaugino/higgsino sector,
the production cross sections of $\tilde{e}_{\rm R}^+\tilde{e}_{\rm R}^-$
and $\tilde{e}_{\rm L}^+ \tilde{e}^-_{\rm R}$ can be exploited to
derive the Yukawa couplings.
However, in the case where the two pairs have almost identical cross sections
and decay modes, the different combinations of
$\tilde{e}_{\rm R}$ and $\tilde{e}_{\rm L}$ can only be distinguished
by the initial beam polarization of the two beams; see Fig.~\ref{fig_susyqn}
(right panels)~\cite{Freitas:2003yp,Moortgat-Pick:2005cw}.\\

\noindent {\bf c) Mass measurements in the continuum}\\
A striking tool at the linear collider are threshold scans, leading, for instance, to
mass measurements of some of the SUSY particles with a precision even 
below the per mil level~\cite{tdr}.
Since threshold scans cost luminosity, it is important to optimize the needed energy steps 
a priori via measurements in the continuum. 
In \cite{uriel,Moortgat-Pick:2005cw} examples have been shown to measure $m_{\tilde{\mu}_{\rm L}}$ and 
$m_{\tilde{\mu}_{\rm R}}$ very accurately up to about 2 per mil in the continuum;  
the dominant $WW$ background has been sufficiently suppressed
only if both beam are polarized: 
the signal-to-background-ratio is about 0.07 (0.45) with $P_{e^-}=+80\%$ 
($P_{e^-}=+80\%$, $P_{e^+}=-80\%$). The polarization of the $e^+$ beams plays a 
lucrative factor in this example, but is rather substantial for the background suppression.

Furthermore, such a precise knowledge of the SUSY particle masses 
is also important to derive the mass of the lightest SUSY particle, the LSP,
very accurately in decay spectra. The LSP presents a promising cold dark matter candidate, 
and therefore a precise knowledge of its properties is crucial.

Such an optimization of threshold scans by accurate continuum measurements is important for all 
linear collider designs, for the ILC as well as for CLIC. Polarized $e^+$ provide an essential
lucrative factor to reach the physics goals in that context.
\\[-.7em]

\noindent {\bf d) Transversely-polarized beams for CP searches}\\
CP-violating phases can be determined via T-odd observables, by 
exploiting spin correlations of the decaying fermions; 
see~\cite{tripel}. 
In \cite{Drees} the use of specific asymmetries with only longitudinally-polarized 
beams for the determination of CP phases has been optimized.
In the case of neutralinos (Majorana fermions), it is, however, even possible to
construct CP-odd asymmetries in the production process 
$e^+e^- \to \tilde{\chi}^0_1 \tilde{\chi}^0_2$ with 
transversely-polarized beams~\cite{Bartl:2005uh}. Both beams have to be polarized. 
In order to measure $A_{\rm CP}$, it is necessary to reconstruct the
directions of the neutralinos. This can be done by analysing the subsequent decays.
This asymmetry can lead to rather high values, even for small phases, i.e.\ 
the parameter range preferred by experimental bounds from the electric dipole moments.
The use of transversely-polarized beams therefore keeps open
an occasionally important  possibility to detect even small CP-violating phases.\\

\begin{figure}
\begin{minipage}{7.2cm}
\caption{
In a R-parity-violating SUSY model, a spin-0 particle
is produced in resonances, the scalar neutrino. The LL polarization configuration points 
directly to the presence of a spin-0 resonance (left plot).
Conversely, a spin-1 resonance, e.g. the $Z^\prime$, is strongest for
the LR configuration (right plot) ~\cite{Moortgat-Pick:2005cw}.
\label{fig_sneu}
}
\end{minipage}
\hspace{1cm}
\begin{minipage}{8.5cm}
\begin{picture}(10,11)
\put(-60,-363){\mbox{\psfig{file=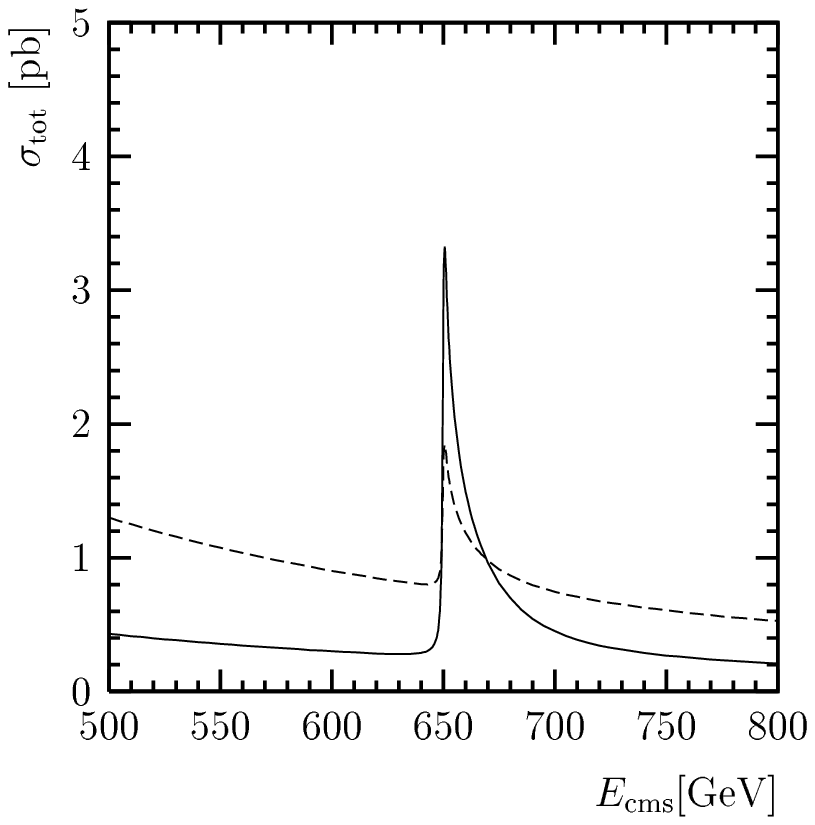,width=1.5\textwidth}}}
\put(10,67){\small $e^+e^-\to \tilde{\nu}_{\tau}\to \mu^+\mu^- $}
\put(58,28){\scriptsize $P_{e^+}={\color{Red}-60\%}$}
\put(58,-15){\scriptsize $P_{e^+}={\color{Olive}+60}\%$}
\put(2,28){\scriptsize $P_{e^-}=-80\%$}
\put(60,-82){\mbox{\includegraphics[height=.03\textheight,width=.22\textheight]{0028-box}}}
\put(58,-72){\tiny $\sqrt{s}$ [GeV]}
\put(+115,-72){\mbox{\psfig{file=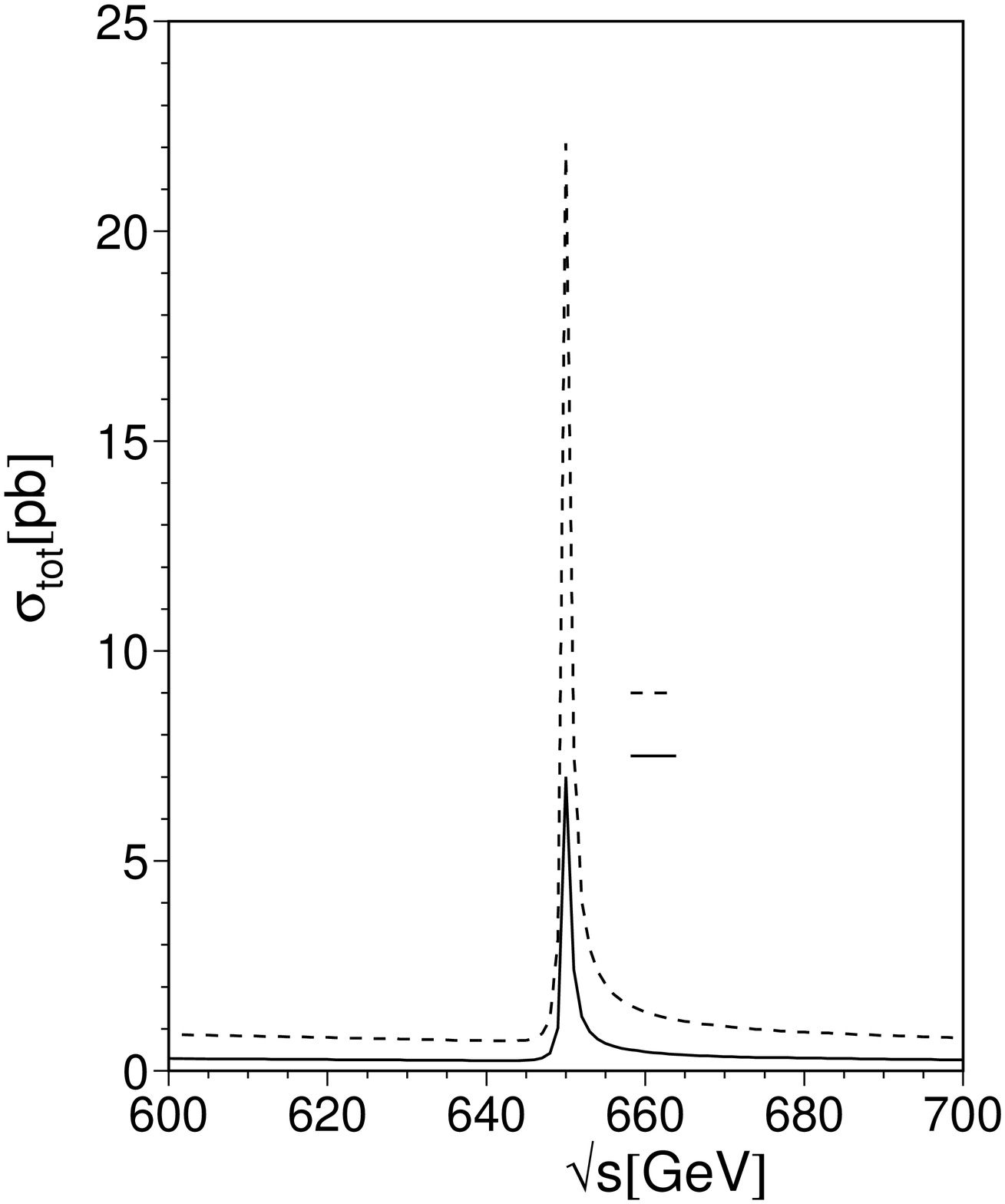,width=.59\textwidth,height=.55\textwidth}}}
\put(150,67){\small $e^+e^-\to Z' \to \mu^+\mu^- $}
\put(200,-25){\mbox{\includegraphics[height=.03\textheight,width=.02\textheight]{0028-box}}}
\put(200,48){\scriptsize $P_{e^+}={\color{Olive}+60\%}$}
\put(200,-25){\scriptsize $P_{e^+}={\color{Red}-60\%}$}
\put(145,48){\scriptsize $P_{e^-}=-80\%$}
\end{picture}
\end{minipage}
\end{figure}

\noindent {\bf e) R-parity-violating SUSY model}\\
The polarization of both beams allows us to
probe directly the spins of particles produced in resonances.
In a R-parity-violating SUSY model a spin-0 particle
is produced in the $s$-channel, the scalar neutrino, with  $\mu^+\mu^-$
in the final state. Since the sneutrino couples only to left-handed $e^\pm$, the peak
is strongest for the LL polarization configuration. Such a signature 
would point directly to the presence of a spin-0 resonance, as in Fig.~\ref{fig_sneu} (left plot).
The SM background is strongly suppressed and one gets a $S/B\sim 11$
for $(P_{e^-},P_{e^+})=(-80\%,-60\%)$, whereas for $(P_{e^-},P_{e^+})=(-80\%,0\%)$ the ratio
is only $S/B\sim 4$~\cite{Spiesberger,Moortgat-Pick:2005cw}.
Conversely, in the case of a spin-1 resonance, e.g. the $Z^\prime$
particle in the SSM model, Fig.~\ref{fig_sneu} (right plot),
the corresponding resonance peak would be strongest for
the LR configuration, with a similar polarization dependence as the SM
background~\cite{Sriemann,Moortgat-Pick:2005cw}. 
This example shows how clearly one could disentangle the
form of the interaction if both beams are polarized.

\vspace{-.5cm}

\hspace*{-.8cm}

\section{Polarized beams in indirect searches}
Some new physics scales, such as those
characterizing gravity in models with extra dimensions or the
compositeness scale of quarks and leptons, could be too large to be
directly accessible at energies of present and future
accelerators. Therefore it will be important
to develop strategies for indirect searches beyond the kinematic limit
for new physics. It is important, however, to get the large model dependence under control.
Thanks to the clear signatures, its high luminosity and beam polarization,
the ILC also has a large discovery potential in indirect searches,
in a largely model-independent approach.\\

\noindent {\bf a) Contact-interaction analysis in Bhabha scattering}\\ 
Effective contact interactions (CI)
represent a general tool for
parametrizing at `low energy' the effects of non-standard dynamics
characterized by exchanges of very high-mass states between the SM particles.\\
 
\begin{figure}[htb]
\setlength{\unitlength}{1cm}
\begin{minipage}{7cm}
\begin{picture}(4.0,6)
\put(0,0){\includegraphics[width=6cm,height=6cm]{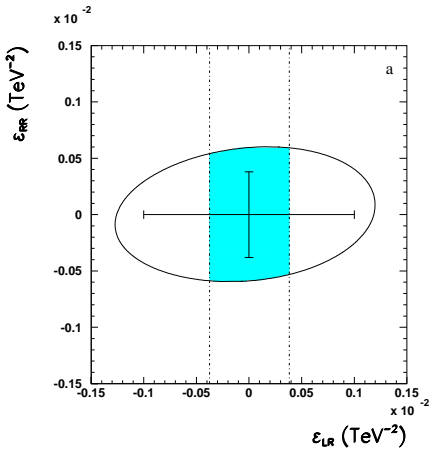}}
\end{picture}
\end{minipage}\hspace{-.3cm}
\begin{minipage}{12cm}
\caption{
In Bhabha scattering the four-fermion CIs are parametrized by three
parameters ($\epsilon_{\rm RR}$, $\epsilon_{\rm LR}$, $\epsilon_{\rm
LL}$). The $t$-channel contributions depend only on $\epsilon_{\rm LR}$, whereas
the $s$-channel contribution depends only on pairs
($\epsilon_{\rm RR}$, $\epsilon_{\rm LR}$), ($\epsilon_{\rm LR}$,
$\epsilon_{\rm LL}$).  In order
to derive model-independent bounds it is necessary to have
both beams polarized. Tight
bounds up to
$5\times 10^{-4}$ TeV$^{-2}$ can be derived via a $\chi^2$ test assuming that no deviations
from the SM
are measured in the observables $\sigma_0$, $A_{\rm FB}$, $A_{\rm LR}$ and $A_{\rm LR,FB}$ (within the
experimental 1~$\sigma$ uncertainty). The study was done at $\sqrt{s}=500$~GeV
~\cite{Pankov:2002qk,Moortgat-Pick:2005cw}.}
\end{minipage}
\end{figure}

\noindent {\bf b) Neutral extra gauge bosons}\\
Extra neutral gauge bosons $Z'$ can be probed by their virtual effects on
cross sections and asymmetries.  For energies below the $Z'$ resonance, measurements of
fermion-pair production are sensitive to the ratio of $Z'$ couplings
and $Z'$ mass. Positron-beam polarization
with $(P_{e^-},P_{e^+})=(80\%,60\%)$  would improve the measurement 
of the $b\bar{b}$ couplings of the $Z'$, even without knowledge of the $Z'$ mass by 
about a factor 1.5, compared with $P_{e^-}=80\%$ only. In the studied example at $\sqrt{s}=500$~GeV
the mass of the $Z'$ was 5 TeV~\cite{LC-TH-2000-006,Moortgat-Pick:2005cw}.
The crucial point is the fact that the systematic
errors can be significantly reduced when both beams are polarized.\\

\begin{figure}
\begin{minipage}{7cm}
\setlength{\unitlength}{1cm}
\begin{picture}(7,5)
\put(0,-.1){\epsfig{file=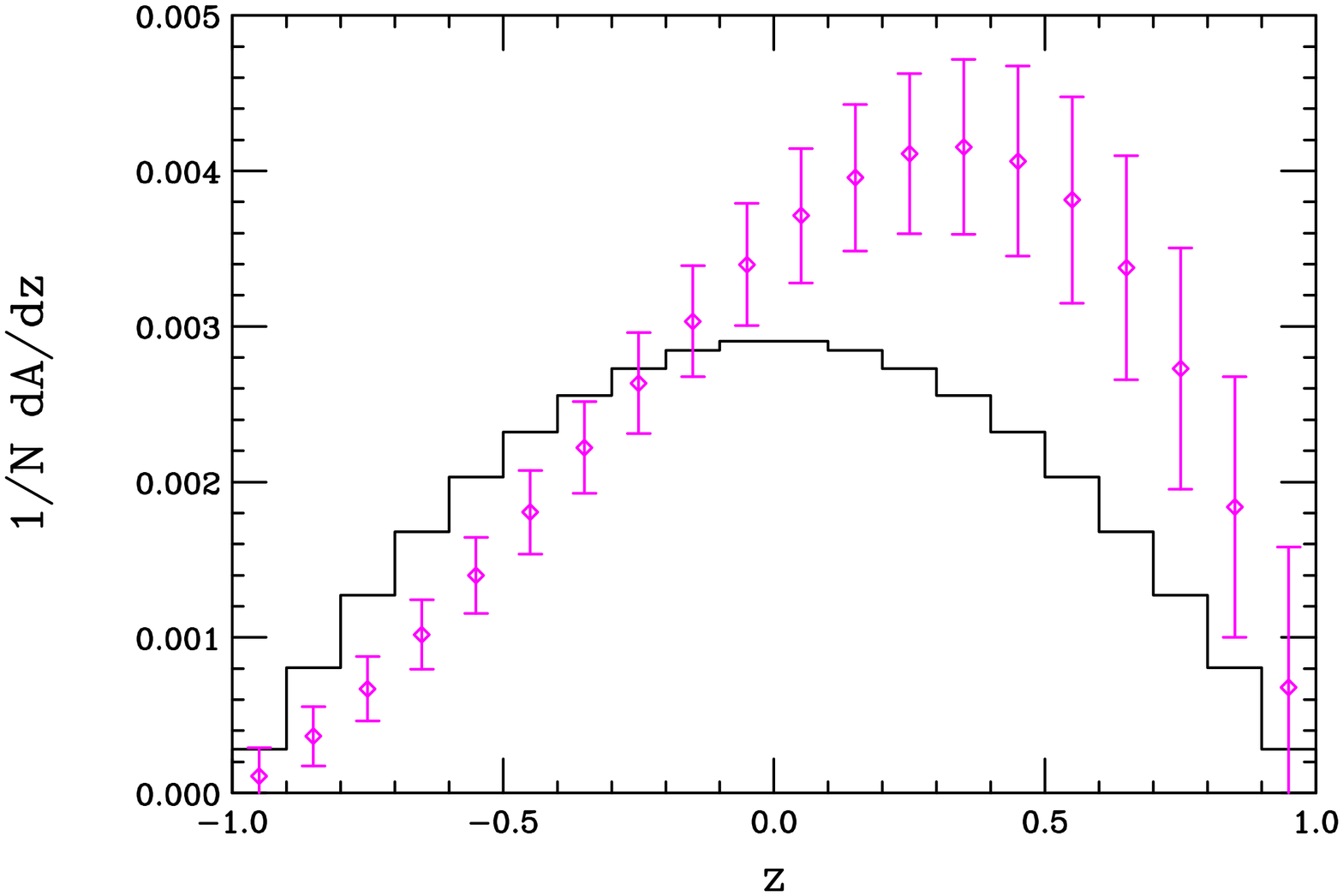,scale=0.32,angle=90}}
\put(0,-.5){\mbox{\includegraphics[height=.03\textheight,width=.22\textheight]{0028-box}}}
\put(0,-.3){\mbox{\includegraphics[width=.015\textheight,height=.22\textheight]{0028-box}}}
\put(3,4.8){\small $e^+e^-\to b$ $\bar{b}$}
\put(2.9,3.6){\small\color{Lila} ADD}
\put(4.1,2.2){\small SM}
\put(3.8,-.1){\small $\cos\theta$}
\put(.3,4.9){\footnotesize $\frac{1}{N}\frac{d A(\phi)}{d \cos\theta}$}
\put(1.3,4.2){\tiny $\sqrt{s}=500$~GeV}
\end{picture}
\end{minipage}\hspace{.5cm}
\begin{minipage}{9.5cm}
\caption{One representative example is the unique distinction between extra dimensions
in the models of Randall-Sundrum (RS) and
Arkani-Hamed, Dimopoulos, Dvali (ADD).
With transversely-polarized beams a new asymmetry in $\sin 2 \phi$ can be constructed.
The new asymmetry vanishes for both the SM and the RS scenario,
so that
a non-zero value unambiguously signals the ADD graviton exchange.
Such a model distinction is achievable up to $\ge 3$~TeV~\cite{Moortgat-Pick:2005cw}.\label{fig_add}}
\end{minipage}
\end{figure}

\noindent {\bf c) Transversely-polarized beams and distinction of graviton models}\\
 Transversely-polarized beams are sensitive to non-standard
interactions, which are not of the current--current type,
such as those mediated by spin-2 gravitons or (pseudo)scalar exchanges,
even in indirect searches. Sensitivities to a high mass scale of, for instance, 
an extra-dimensional model 
are achievable. Even different models with large extra dimensions can be
distinguished; see Fig.~\ref{fig_add}~\cite{Rizzo:2002ww,Moortgat-Pick:2005cw}. 
The study was done for $\sqrt{s}=500$~GeV.
Success in identifying new physics even in indirect
searches using polarized $e^-$ and $e^+$ beams would represent a big step forward
for our understanding of fundamental interactions. Both beams have to be polarized to observe
effects from transversely-polarized beams at the linear collider.

\section{Polarized beams in high-precision measurements at GigaZ}
\begin{figure}[htb]
\begin{minipage}{9.6cm}
\caption{ With the polarization of both beams, using the Blondel scheme,
assuming 80\% polarization for
electrons and 60\% for positrons, an accuracy of
$\Delta \sin^2\theta_{\rm eff} = 1.3 \times 10^{-5}$ can be
achieved in the leptonic final state~\cite{tdr}.
A polarization degree of $P_{e^+}\sim 60\%$ is sufficient,
assuming  $\Delta P_{e^-}/P_{e^-}=\Delta P_{e^+}/P_{e^+}=0.5\%$.
\label{fig_gigaz_asy}
}
\end{minipage}
\hspace{1.cm}
\begin{minipage}{4.5cm}
\begin{picture}(6,12)
\setlength{\unitlength}{1cm}
\put(-.5,-3){\mbox{\includegraphics[height=.30\textheight]{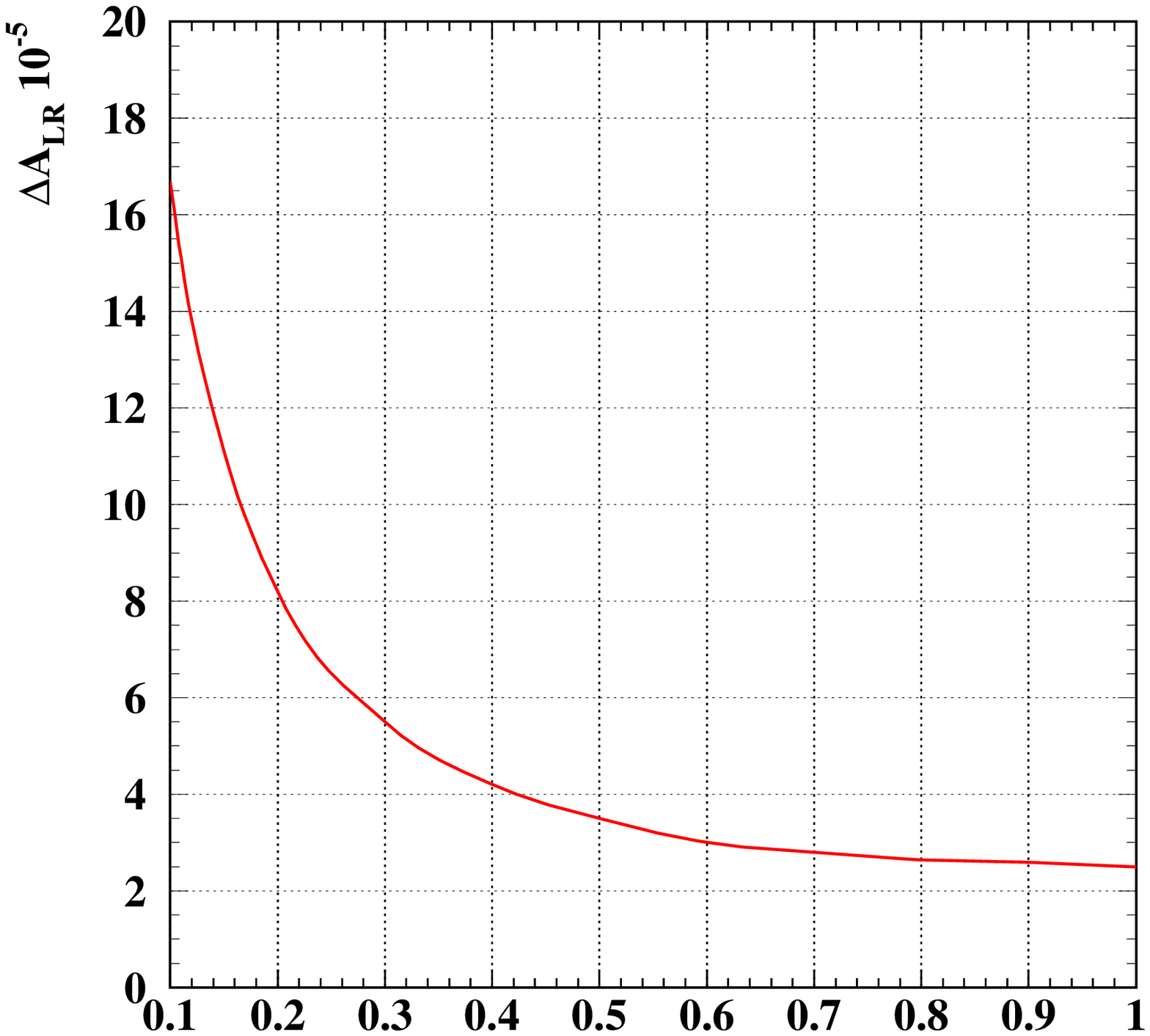}}}
\put(2,.5){\mbox{\includegraphics[height=.07\textheight,width=.15\textheight]{0028-box}}}
\put(3.,-1.2){\mbox{\includegraphics[height=.078\textheight,width=.09\textheight]{0028-box}}}
\put(.8,2.4){\mbox{\includegraphics[height=.02\textheight,width=.05\textheight]{0028-box}}}
\put(3.2,-.5){\boldmath{$\Bigg \Uparrow$}}
\put(.7,2.6){\tiny $P_{e^-}=80\%$}
\put(3.7,-.1){\tiny leads to}
\put(3.7,-.5){\tiny $P_{e^+}=60\%$}
\put(2.1,1.1){{\tiny
\renewcommand{\arraystretch}{1.5}
\begin{tabular}{cc}
$\Delta\sin^2\theta_{\rm eff}$ & $(P_{e^-},P_{e^+})$ \\
$9.5\cdot 10^{-5}$ & (80\%,0)\\
{\color{Red}$1.3\cdot 10^{-5}$} & (80\%,60\%)\\
\end{tabular}
}}
\put(5.2,-2.9){\small $P_{e^+}$}
\end{picture}
\end{minipage}
\end{figure}

Extremely sensitive tests of the SM can be performed with the help of
electroweak precision observables. These can be measured with very high
accuracy at the GigaZ option of the ILC, i.e. running with high luminosity
at the $Z$-boson resonance. Measuring accurately the
left--right asymmetry allows a determination of the effective weak mixing angle
 $\sin^2\theta_{\rm eff}$ with the highest precision.  However, in order to
exploit the gain in statistics at GigaZ, the relative uncertainties on the beam
polarization have to be kept below 0.1\%. This ultimate precision cannot be
reached with Compton polarimetry, but by using a modified Blondel scheme,
which requires the polarization of both beams; see 
Fig.~\ref{fig_gigaz_asy}~\cite{Hawkings:1999ac,Moortgat-Pick:2005cw}.
So far the GigaZ option is only discussed as a later upgrade 
for the ILC. But physics arguments could require that a quick and cheap upgrade path to GigaZ
be provided straight after the $\sqrt{s}=500$~GeV stage.

\begin{figure}[htb]
\begin{minipage}{7.cm}
\setlength{\unitlength}{1cm}
\begin{picture}(7,7)
\put(0.,0)
{\mbox{\epsfig{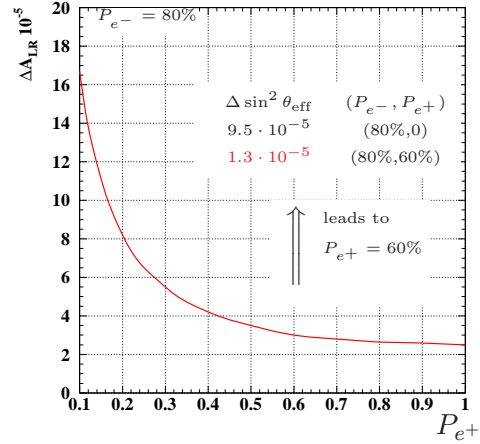}}}
\end{picture}
\end{minipage}
\begin{minipage}{10cm}
\caption{ \label{fig_gigaz_mh}
The theoretical predictions for $\sin^2 \theta_{\rm eff}$ in terms
of $m_h$, the mass of the Higgs boson in the SM or
the mass of the lightest Higgs boson in the MSSM,
respectively, are compared with the experimental
accuracies obtainable at GigaZ~\cite{Moortgat-Pick:2005cw}.
The bounds on $m_h$ in the SM are improved by about 1 order of magnitude with
$(|P_{e^-}|,|P_{e^+}|)=(80\%,60\%)$ instead of $(|P_{e^-}|,|P_{e^+}|)=(80\%,0\%)$; 
}
\end{minipage}
\end{figure}

\begin{figure}[htb]
\begin{minipage}{10cm}
\caption{ \label{fig_gigaz_m12}
The precision measurement of $\sin^2 \theta_{\rm eff}$
yields constraints on the allowed range for the SUSY
mass parameter $m_{1/2}$ in a specific model, the
CMSSM. The allowed range of $m_{1/2}$ is reduced by a 
factor of about 5 when using
$(|P_{e^-}|,|P_{e^+}|)=(80\%,60\%)$ 
instead of $(|P_{e^-}|,|P_{e^+}|)=(80\%,0\%)$.
Experimental constraints from LEP
searches and cold-dark-matter searches have been
taken into account~\cite{Moortgat-Pick:2005cw}.
}
\end{minipage}
\begin{minipage}{7.5cm}
\setlength{\unitlength}{1cm}
\begin{picture}(7,4.5)
\put(0.5,-.5)
{\mbox{\epsfig{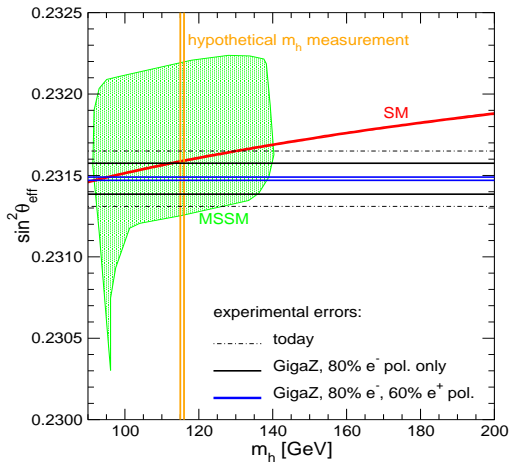}}}
\end{picture}
\end{minipage}
\end{figure}

Because of the gain of about 1 order of magnitude in the accuracy of $\sin^2 \theta_{\rm eff}$,
the bounds on $m_h$ in the SM improve by about 1 order of magnitude (see Fig.~\ref{fig_gigaz_mh}), and
the allowed range of $m_{1/2}$ is reduced by a factor of about 5 when using
$(|P_{e^-}|,|P_{e^+}|)=(80\%,60\%)$ instead of $(|P_{e^-}|,|P_{e^+}|)=(80\%,0\%)$; 
see Fig.~\ref{fig_gigaz_m12}~\cite{Svenni-Georg,Moortgat-Pick:2005cw}.
Such a piece of information could become essential to outline the needed energy scale for future
linear-collider options and it will be of great importance to get the most accurate information from 
the GigaZ option.

\vspace{.5cm}\hspace{-.8cm}


\section{Conclusions}
It is generally agreed that the clean and precise
environment of $e^+e^-$ collisions at the ILC is ideally suited to the search
for new physics and for determining precisely the underlying structure of
the new interactions. This physics case 
is independent of the results of the LHC. The results 
of both the LHC and the ILC
will be crucial to assess 
the physics programme in the multi-TeV energy region of the CLIC design.
The physics potential of both the ILC and CLIC will 
greatly benefit from the availability of 
polarized $e^-$ and $e^+$ beams.

Polarization of both beams at the ILC would be ideal for
facing both expected and unforeseen challenges in physics analyses, for instance
to fix the chirality of the couplings and to enable the higher precision for
the polarization measurement itself as well as for polarization-dependent observables.
It  provides a powerful tool for studying SM physics as well as new physics,
such as precisely analyzing the top and Higgs properties,
discovering new particles, analyzing signals model-independently
and resolving precisely the underlying model.
We have demonstrated that the full potential of the ILC will be realized
only with a polarized positron beam together with a polarized 
electron beam~\cite{Moortgat-Pick:2005cw}.
That is in particular relevant for the first energy stage of the ILC at 
$\sqrt{s}=500$~GeV, since otherwise 
only limited experimental information on the new physics might be available. 

Polarized $e^+$ in addition to polarized $e^-$ lead to   
substantial improvements and/or lucrative statistical enhancements
in the physics analyses. In some cases the lucrative character becomes
also substantial to optimize the outcome at the ILC at $\sqrt{s}=500$~GeV. 
For instance, in many top analyses
a factor 3 can be gained when using $(|P_{e^-}|,|P_{e^+}|)=(80\%,60\%)$ 
instead of $(|P_{e^-}|,|P_{e^+}|)=(80\%,0\%)$. Also challenging Higgs studies, for 
instance the analyses of the top Yukawa couplings and the triple Higgs couplings, 
that are of great importance for the understanding of the electroweak symmetry breaking,
benefit by a factor up to 2.5 when using $(|P_{e^-}|,|P_{e^+}|)=(80\%,60\%)$ 
instead of $(|P_{e^-}|,|P_{e^+}|)=(80\%,0\%)$.

Having two polarized beams available is crucial for uniquely
determining the properties and the quantum numbers of new particles,
and for testing fundamental model assumptions, as we have demonstrated in the
specific example of supersymmetry. The larger number of observables
accessible with two polarized beams provides better tools for
revealing the structure of the underlying physics, determining new
physics parameters, getting background processes and systematic
uncertainties under control and enabling model-independent analyses.

Furthermore, with both beams polarized, one has the possibility to exploit
transversely-polarized beams for physics studies. This option provides
new and efficient observables for the detection of possible sources of
CP violation. Additionally, it is a unique tool for distinguishing between
different models with extra spatial dimensions, far below the
threshold of the spin-2 excitations.

To fully exploit high-precision tests of the Standard Model at GigaZ,
both beams must be polarized. The measurement of the electroweak
precision observables is of utmost importance to test the SM and to
derive precise bounds on the Higgs mass as well as to determine 
possible parameter
ranges of new physics as, for instance, in supersymmetry.
Improvements up to an order of magnitude can be obtained when using
$(|P_{e^-}|,|P_{e^+}|)=(80\%,60\%)$ instead of
$(|P_{e^-}|,|P_{e^+}|)=(80\%,0\%)$.  Such precise predictions from
indirect searches either at $\sqrt{s}=500$~GeV or at GigaZ will be
crucial to outline  the physics programme of future linear collider
options and to determine the physics potential of possible 
high-energy designs.



\end{document}